\documentclass[12pt]{article}

\begin{document}

\begin{center}

{\Large {Kallen-Lehmann representation for Spinor-Scalar Loops in de Sitter Space-time. Spectral equations}}

\vspace{1,5cm}

{Boris L. Altshuler}\footnote{E-mail addresses: baltshuler@yandex.ru $\,\,\,  \& \,\,\,$  altshul@lpi.ru}

\vspace{1cm}

{\it Theoretical Physics Department, P.N. Lebedev Physical
Institute, \\  53 Leninsky Prospect, Moscow, 119991, Russia}

\vspace{1,5cm}

\end{center}

{\bf Abstract:} Compact formulas (1), (28) presented in this paper permit to formulate the Kallen-Lehmann harmonic decompositions of the 
products of two spinor and of spinor and scalar  harmonic functions on AdS and Wightman functions on dS spaces; the correct flat space 
limit of the corresponding Kallen-Lehmann densities is demonstrated, the connection between poles of the Kallen-Lehmann density and the late-time divergence of the spinor-scalar loop is traced in the Yukawa model with light scalar field.Also in Sec. 4  the relatively simple, suitable for calculations Kallen-Lehmann representations of the real parts of spinor and scalar one-loop self-energies on dS space-time are proposed, and corresponding spectral equations for conformal dimensions in the chain approximation are written down.

\vspace{0,5cm}


\newpage

\section{Introduction}

\quad  According to rhe astrophysical observations, during the inflation period in the first moments after the Big Bang and in the era of accelerated 
expansion observed today, the metric of our universe may be approximately described as metric of de Sitter space-time. The study of quantum fields 
against de Sitter or quasi-de Sitter background has become an important direction of work. Here we concentrate on the involving Dirac 
spin $1/2$ field one-loop corrections to bulk propagators on de Sitter space-time (dS).
The tree diagrams of spinor fields on dS as well as on anti-de Sitter (AdS) spaces are considered in \cite{Schaub} where 
there are references to both (dS and AdS) lines of such studies and important accompanying expressions are presented. 
Fermion loops on the dS background were studied in \cite{Ferm0511140} - \cite{Ferm2503} but the technique of the present article essentially differes 
from that of these works.

The works on quantum fields of zero and integer spin on de Sitter space-time were also an essential background of the present 
research. First of all those are pioneer papers \cite{Bros91} - \cite{Hollands}, as well as papers considering dS propagators and correlators in the context of the relationship of dS and AdS \cite{Sleight1906} - \cite{Pened2505}.
Also papers \cite{Boyan1103} - \cite{Miao2405} (the list is obviously incomplete) dedicated to the possible loops-induced IR instability of de Sitter 
space-time proved to be of great importance.

In this paper the Kallen-Lehmann representation for the one-loop self-energy diagrams of scalar and spinor fields formed correspondingly 
by two spinors and by spinor and scalar are derived in the model with Yukawa interaction $L_{int} = e {\bar\psi}\psi \phi$ on de Sitter 
space-time of $D = d + 1$ dimensions. The article continues and develops a similar studies on the anti-de Sitter space \cite{alt-1}, \cite{alt-2}.

The paper begins with a description of previously known and partly new results concerning one-loop diagrams 
with spinors on the background of the Euclidean anti-de Sitter (EAdS) space; the Kallen-Lehmann decompositions of spinor-spinor and spinor-scalar EAdS 
"harmonic bubbles" (products of scalar and spinor AdS harmonic functions) are derived. 

These expressions are used in Section 3 where similar Kallen-Lehmann decompositions of the products of scalar and spinor Wightman functions 
on dS background are presented. The obtained expressions for the Kallen-Lehmann densities have the correct flat space limit to the corresponding densities of the Kallen-Lehmann representations of spinor and scalar self-energies in the flat space Yukawa model. In the Yukawa model with a light scalar field, the connection between the poles of the KL density obtained in this work and the known for light quantum scalar fields IR-divergence near the future dS horizon is traced.

In Section 4 two-scalars, two-spinors ans spinor-scalar one-loop self-energy diagrams formed by the Feynman time-ordered Green functions are studied.
It is proved that Kallen-Lehmann representation in a form of the principal value of the corresponding spectral integral is valid for the real part of the two-scalars self-energy on dS. In general the hypothesis is put forward that any one-loop self-energy diagram on de Sitter space-time can be expressed as the Kallen-Lehmann spectral integral similar in its structure to the corresponding one in Minkowski space-time. Based on this hypothesis, spectral equations for the masses of spinor and scalar bound states are written down in the "chain" approximation of the corresponding Dyson equations.

In Sections 3, 4 dedicated to de Sitter space-time the well known Kallen-Lehmann expressions for heavy scalar fields ($M/H > h$, $H = L_{dS}^{-1}$ is Hubble constant, $h = (D-1)/2 = d/2$) of the principa-lvalue series are written down, like it is done in \cite{Eps06} - \cite{Eps12}, \cite{Gorb2108}, \cite{Pened2107}, \cite{Eps24-2}, and following the named works we omit in the corresponding KL expressions the complementary series pole terms when masses of scalar fields are analytically continued to the region of light masses $0 < M/H < h$. These pole terms are calculated in \cite{Eps0901} (App. C), \cite{Pened2306}, \cite{UGO1701}, their origin is perhaps most clearly seen in $S^{D} \to dS^{D}$ analytical continuation when Kallen-Lehmann representation on dS is obtained from the Watson-Zommerfeld one on $S^{D}$, see e.g. \cite{Marolf}, \cite{Hollands}, \cite{Gorb2108}, \cite{Pened2107}. For us here it is important that the long-distance IR subtleties characteristic of light or massless scalar fields are reflected in the principle-value Kallen-Lehmann expressions continued analytically into the region of small scalar fields' masses, which is demonstrated by a specific example at the end of Section 3.

For spinor fields on de Sitter space there is no separation into principal and complementary series. The question of whether it is necessary to supplement the spinor-scalar Kallen-Lehmann expressions obtained in the present paper with some pole terms when light scalar fields are present in the loop diagrams of the Yukawa model requires special study and is beyond the scope of this work.

The Conclusion outlines the possible areas of application of the obtained results and ways of their further development.

\section{Kallen-Lehmann decomposition of product of AdS spinor-scalar harmonic functions}

\qquad The starting point is the calculation in \cite{alt-1} of the spinor-including EAdS one-loop cntributions to the spinor-spinor and scalar-scalar 
correlators where they were expressed through the one and the same compact function of conformal dimensions of two spinor 
fields $\psi(x)$, $\chi(x)$ and scalar field $\phi(x)$:

\begin{eqnarray}
\label{1}
{\cal {\bf A}}(\Delta_{\psi}, \Delta_{\chi}; \Delta_{\phi}) =  \frac{\Gamma\left(\frac{h}{2} \pm \alpha \right) \, \Gamma\left(\frac{h}{2} \pm \beta \right) \,\Gamma\left(\frac{1}{2} + \frac{h}{2} \pm \gamma\right) \, \Gamma\left(\frac{1}{2} + \frac{h}{2} \pm \delta \right)}{\Gamma\left(\frac{1}{2} \pm \nu_{\psi}\right) \, \Gamma\left(\frac{1}{2} \pm \nu_{\chi}\right) \, \Gamma( \pm \nu_{\phi})}, \nonumber
\\
\\
\alpha = \frac{\nu_{\phi}  + \nu_{\psi} - \nu_{\chi}}{2}, \, \beta = \frac{\nu_{\phi}  - \nu_{\psi} + \nu_{\chi}}{2}, \, \gamma = \frac{\nu_{\psi}  + \nu_{\chi} +\nu_{\phi}}{2}, \, \delta = \frac{\nu_{\psi}  + \nu_{\chi} - \nu_{\phi}}{2},  \nonumber
\end{eqnarray}
here

\begin{eqnarray}
\label{2}
 h = \frac{d}{2}, \quad \Gamma(a \pm b) = \Gamma(a+b) \Gamma(a - b), \quad \Delta_{\psi,\chi} = h + \nu_{\psi,\chi} \nonumber
 \\
 \\
 \Delta_{\phi} = h + \nu_{\phi}, \quad \nu_{\psi,\chi} = m_{\psi, \chi} L_{AdS}, \quad \nu_{\phi} = \sqrt{h^{2} + M^{2} L^{2}_{AdS}}, \nonumber
\end{eqnarray}
$m_{\psi}, m_{\chi}, M$ are masses of spinor and scalar fields, $L_{AdS}$ is radius of curvature of AdS space.

Function ${\cal {\bf A}}(\Delta_{\psi}, \Delta_{\chi}; \Delta_{\phi})$ (\ref{1}) is actually the EAdS "scalar - two spinors" twin of the 
similar "three scalars" function introduced in seminal paper \cite{pened1011}:

\begin{eqnarray}
\label{3}
 \Theta (\nu_{1},  \nu_{2}, \nu_{3}) = \frac{\Pi_{\{\sigma_{i} = \pm\}} \Gamma\left( \frac{h + \sigma_{1}\nu_{1} + \sigma_{2}\nu_{2} + \sigma_{3}\nu_{3} }{2}\right)}{\Pi_{i = 1}^{3}\Gamma(\nu_{i}) \Gamma(-\nu_{i})} =  \nonumber
 \\
 \\
 = \frac{16 \pi^{2h} \Gamma(2h)} {L_{AdS}^{d + 1}} \cdot \int_{AdS} dX \Omega^{(0)}_{\nu_{1}}(X,Y) \, \Omega^{(0)}_{\nu_{2}}(X,Y) \, \Omega^{(0)}_{\nu_{3}}(X,Y),  \nonumber 
\end{eqnarray} 
$\Omega^{(0)}_{\nu}(X,Y)$ is the scalar field harmonic function in EAdS. Information on scalar and spinor EAdS harmonic functions and their properties may be found in \cite{Schaub}, \cite{alt-1} - \cite{pened1404}. It is worthwhile to note that a formula just as beautiful as (\ref{3}), although slightly different in the denominator, was obtained in dS in \cite{Eps0901}; de Sitter space will be discussed in the next Section.

In the same way function ${\cal {\bf A}}$ (\ref{1}) is proportional to the integrated over EAdS product of two spinor and one scalar harmonic function:

\begin{equation}
\label{4}
{\cal {\bf A}}(\Delta_{\psi}, \Delta_{\chi}; \Delta_{\phi}) = \frac{2^{5} \pi^{2h + 2} \Gamma(2h)}{L_{AdS}^{d + 1} \, {\it dim \gamma}} \cdot \int_{AdS} dX {\rm Tr} [ \Omega^{(\frac{1}{2})}_{\nu_{\psi}}(X,Y) \, \Omega^{(\frac{1}{2})}_{\nu_{\chi}}(X,Y)] \, \Omega^{(0)}_{\nu_{\phi}}(X,Y),
\end{equation}
here trace is taken over spinor indices, ${\it dim \gamma} = {\rm Tr} {\hat 1} = 2^{[D/2]}$, $[D/2]$ is the integer part of half of dimensionality $D = d + 1$ of space.

Functions ${\cal {\bf A}}$ (\ref{1}), (\ref{4}) and $\Theta$ (\ref{3}) play key role in calculation of the Kallen-Lehmann decomposition of the corresponding products of two harmonic functions. For "two scalars harmonic bubble" this decomposition looks as \cite{pened1011}

\begin{equation}
\label{5}
\Omega^{(0)}_{\nu_{1}}(X,Y) \, \Omega^{(0)}_{\nu_{2}}(X,Y) = \int^{+\infty}_{- \infty} dc \, B^{(0,0)}_{\nu_{1}, \nu_{2}} (c) \, \Omega^{(0)}_{c}(X,Y),
\end{equation}
where

\begin{eqnarray}
\label{6}
B^{(0,0)}_{\nu_{1}, \nu_{2}} (\nu_{3}) = \frac{\int_{AdS} dX \Omega^{(0)}_{\nu_{1}}(X,Y) \, \Omega^{(0)}_{\nu_{2}}(X,Y) \, \Omega^{(0)}_{\nu_{3}}(X,Y)}{L_{AdS}^{d + 1} \, \, \, \Omega^{(0)}_{\nu_{3}}(X,X)} = \qquad \nonumber
\\
\\
= \frac{1}{8 \pi^{h} \Gamma (h)} \, \, \, \, \Theta (\nu_{1},  \nu_{2}, \nu_{3}) \cdot \frac{ \Gamma(\pm \nu_{3})}{\Gamma (h \pm \nu_{3})}. \qquad \qquad  \qquad  \nonumber
\end{eqnarray}
Here

\begin{equation}
\label{7}
\Omega^{(0)}_{\nu}(X,X) = \frac{\Gamma(h)} { 2 \pi^{h} \Gamma(2h)} \, \frac{\Gamma (h \pm \nu)}{\Gamma (\pm \nu)}.
\end{equation}

As it is shown in Appendix $D$ of \cite{pened1011}, the first line of (\ref{6}) immediately follows if we multiply left and right sides of (\ref{5}) by $\Omega^{(0)}_{\nu_{3}}(Y,Z)$, integrate over $Y$, use the convolution formula

\begin{equation}
\label{8}
\int_{AdS} dY \Omega^{(0)}_{c}(X,Y) \, \Omega^{(0)}_{\nu_{3}}(Y,Z) = L_{AdS}^{d + 1} \, \frac{1}{2}\, [\delta(c - \nu_{3}) + \delta(c + \nu_{3})] \, \Omega^{(0)}_{\nu_{3}}(X,Z),
\end{equation}
and finally put $Z = X$. We note that density $B^{(0,0)}_{\nu_{1}, \nu_{2}} (c)$ in decomposiyion (\ref{5}) as well as $\Omega^{(0)}_{c}(X,Y)$ are both symmetric under the change of sign of $c$, thus integral over $c$ in (\ref{5}) may be taken from $0$ to $+\infty$.

Exactly the same proceedure with use of the convolution formula for spinor harmonic functions \cite{Schaub}

\begin{equation}
\label{9}
\int_{AdS} dY \Omega^{(\frac{1}{2})}_{\nu_{1}}(X,Y) \, \Omega^{(\frac{1}{2})}_{\nu_{2}}(Y,Z) = L_{AdS}^{d + 1} \, \frac{1}{2}\, [\delta(\nu_{1} - \nu_{2}) + \delta(\nu_{1} + \nu_{2})] \, \Omega^{(\frac{1}{2})}_{\nu_{1}}(X,Z)
\end{equation}
may be applied to EAdS harmonic bubbles with spinors. Namely for the Kallen-Lehmann decomposition of spinor-scalar harmonic bubble it is obtained:

\begin{equation}
\label{10}
\Omega^{(\frac{1}{2})}_{\nu_{\chi}}(X,Y) \, \Omega^{(0)}_{\nu_{\phi}}(X,Y) = \int^{+\infty}_{- \infty} dc \, B^{(\frac{1}{2},0)}_{\nu_{\chi}, \nu_{\phi}} (c) \, \Omega^{(\frac{1}{2})}_{c}(X,Y),
\end{equation}
where

\begin{eqnarray}
\label{11}
B^{(\frac{1}{2},0)}_{\nu_{\chi}, \nu_{\phi}} (\nu_{\psi}) = \frac{ \int_{AdS} dX {\rm Tr} [ \Omega^{(\frac{1}{2})}_{\nu_{\psi}}(X,Y) \, \Omega^{(\frac{1}{2})}_{\nu_{\chi}}(X,Y)] \, \Omega^{(0)}_{\nu_{\phi}}(X,Y),}{L_{AdS}^{d + 1} \quad {\rm Tr}[\Omega^{(\frac{1}{2})}_{\nu_{\psi}}(X,X)]} = \qquad \nonumber
\\
\\
= \frac{1}{16 \pi^{h + 2} \Gamma (h)} \,\, {\cal {\bf A}}(\nu_{\psi}, \nu_{\chi}; \nu_{\phi}) \cdot \frac{\Gamma(\frac{1}{2} \pm \nu_{\psi})}{\Gamma (h + \frac{1}{2}\pm \nu_{\psi})}. \qquad \qquad  \qquad  \nonumber
\end{eqnarray}
Here, see \cite{Allais}, \cite{Aros},

\begin{equation}
\label{12}
{\rm Tr}[ \Omega^{(\frac{1}{2})}_{\nu_{\psi}}(X,X)] = {\it dim \gamma} \, \, \frac{\Gamma(h)} { 2 \pi^{h} \Gamma(2h)} \, \frac{\Gamma (h + \frac{1}{2}\pm \nu_{\psi})}{\Gamma (\frac{1}{2}\pm \nu_{\psi})}.
\end{equation}
Now, as can be seen from (\ref{11}) and (\ref{1}), density $B^{(\frac{1}{2},0)}_{\nu_{\chi}, \nu_{\phi}} (c)$ is not symmetric under the change of sign of $c$; symmetric and antisymmetric parts of $B^{(\frac{1}{2},0)}_{\nu_{\chi}, \nu_{\phi}} (c)$ correspond to two Kallen-Lehmann densities typical in the spectral representation of spinor Green functions, see discussion in the next Section.

In the same way for the Kallen-Lehmann decomposition of the EAdS two-spinors harmonic bubble it  is obtained

\begin{equation}
\label{13}
{\rm Tr} [\Omega^{(\frac{1}{2})}_{\nu_{\psi}}(X,Y) \, \Omega^{(\frac{1}{2})}_{- \nu_{\chi}}(X,Y)] = \int^{+\infty}_{- \infty} dc \, B^{(\frac{1}{2},\frac{1}{2})}_{\nu_{\psi}, \nu_{\chi}} (c) \, \Omega^{(0)}_{c}(X,Y),
\end{equation}
where

\begin{eqnarray}
\label{14}
B^{(\frac{1}{2},\frac{1}{2})}_{\nu_{\psi}, \nu_{\chi}} (\nu_{\phi}) = \frac{ \int_{AdS} dX {\rm Tr} [ \Omega^{(\frac{1}{2})}_{\nu_{\psi}}(X,Y) \, \Omega^{(\frac{1}{2})}_{- \nu_{\chi}}(X,Y)] \, \Omega^{(0)}_{\nu_{\phi}}(X,Y)}{L_{AdS}^{d + 1} \quad \Omega^{(0)}_{\nu_{\phi}}(X,X)} = \qquad \nonumber
\\
\\
= \frac{{\it dim \gamma}}{16 \pi^{h + 2} \Gamma (h)} \,\, {\cal {\bf A}}(\nu_{\psi}, - \nu_{\chi}; \nu_{\phi}) \cdot \frac{\Gamma(\pm \nu_{\phi})}{\Gamma (h \pm \nu_{\phi})}. \qquad \qquad  \qquad  \nonumber
\end{eqnarray}
Change of sign of $\nu_{\chi}$ correspondes to the change of $\Delta_{\chi}$ to $d - \Delta_{\chi}$ in the arguments of ${\cal {\bf A}}$ in 
the LHS of (\ref{1}) and physically means the change of sign of mass of field $\chi (x)$ because two-fermion loop is formed  by particle and antiparticle, of equal masses as a rule.

All expressions for the Kallen-Lehmann decompositions of the EAdS harmonic bubbles described above will be used in the next Section dedicated to the quantum loops with spinors on de Sitter space-time. 

The method for obtaining these decompositions looks quite simple, but this simplicity is deceptive. The non-trivial task is to derive expressions (\ref{3}), (\ref{4}) for three harmonic functions integrated over EAdS space. Formula (\ref{3}) for three scalar harmonic functions was directly derived in \cite{pened1011} whereas the form (\ref{1}) of function {\cal {\bf A}} in (\ref{4}) was obtained in \cite{alt-1} in more simple way - through calculation of the corresponding correlators. Unfortunately approach of \cite{alt-1} does not permit to determine the overall constant in (\ref{4}) due to the universal conformal divergence in the calculation of conformal correlators \cite{Giombi1}.

The proportionality coefficient in (\ref{4}) is determined in this work on the physical grounds - from the requirement that the Kallen-Lehmann 
representations of the loops with spinors in de Sitter space-time obtained in the next section have the form known from textbooks in the limit 
of flat Minkowski space-time. 

\section{Kallen-Lehmann decomposition of product of dS spinor-scalar Wightman functions}

\qquad The idea to use the well-developed apparatus in Euclidean anti-de Sitter space to obtain the results in de Sitter space-time was proposed in \cite{Sleight1906}, developed and applied in \cite{Taron1907} - \cite{Gorb2108}, \cite{Pened2306}. The way to transform EAdS expressions to dS ones is the analytical continuation $L_{AdS} \to \pm i L_{dS}$, $z \to \pm i \eta$ where $\pm$ correspond to the left and right parts of Keldysh contour in dS, $z$ is radial coordinate of the EAdS metric in Poincare patch and $\eta$ is conformal time of dS metric of the most plus signature:

\begin{equation}
\label{15}
ds_{AdS}^{2} = L_{AdS}^{2} \frac{dz^{2} + {\vec{dx}}^{2}}{z^{2}}  \to ds_{dS}^{2} = L_{dS}^{2} \frac{- d\eta^{2} + {\vec{dx}}^{2}}{\eta^{2}},
\end{equation}
at $L_{dS} \to \infty$ metric $ds_{dS}^{2}$ transforms into the Minkowski metric of the $d +1$ dimensional flat space-time.

Then conformal EAdS indices defined in (\ref{2}) become imaginary (for scalar field this is true for the principal values of field's mass $M L_{dS} > h$):

\begin{equation}
\label{16}
\nu_{\psi, \chi} \to i \lambda_{\psi, \chi} = i m_{\psi, \chi} L, \quad \nu_{\phi} \to i \lambda_{\phi} = i \sqrt{M^{2} L^{2} - h^{2}}.
\end{equation}
From now on we omit the subscribt $dS$ at $L_{dS}$ that is change $L_{dS} \to L$.

The role of harmonic functions in EAdS is played in dS by Wightman functions $W^{0}(X, Y) = \langle 0 |\phi(X) \phi(Y) | 0 \rangle $, and the 
same for spinor fields, here $| 0 \rangle$ is the Bunch-Davies vacuum. Wightman functions and corresponding EAdS harmonic functions are related to each 
other in a simple way, see \cite{Sleight1906}, \cite{Taron2109}, \cite{Gorb2108} for scalars and \cite{Schaub} for spinors:

\begin{eqnarray}
\label{17}
W^{(0)}_{\lambda_{\phi}} (X, Y) = \Gamma (\pm i \lambda_{\phi}) \cdot \Omega^{(0)}_{i\lambda_{\phi}}(X, Y); \nonumber
\\
\\
W^{(\frac{1}{2})}_{\lambda_{\psi}} (X, Y) = \Gamma (\frac{1}{2} \pm i \lambda_{\psi}) \cdot \Omega^{(\frac{1}{2})}_{i\lambda_{\psi}}(X, Y). \nonumber
\end{eqnarray}
The first of relations (\ref{17}) is also valid for the light scalar fields of the complementary series -
with replacement of imaginary arguments of Gamma functions to real ones (see (\ref{16}) for $ML < h$).
 
These formulas are what we need now. Changing arguments in EAdS convolution expression (\ref{8}) according to (\ref{16}) ($c \to ic$, $\nu_{3} \to i \lambda_{3}$) and multiplying (\ref{8}) by $\Gamma (\pm i c) \cdot \Gamma (\pm i \lambda_{3})$ we get the convolution formula for scalar Wightman functions:

\begin{equation}
\label{18}
\int_{dS} dY W^{(0)}_{c}(X,Y) \, W^{(0)}_{\lambda_{3}}(Y,Z) = L^{d + 1} \, \Gamma (\pm i \lambda_{3}) \, \delta(c - \lambda_{3}) \, W^{(0)}_{\lambda_{3}}(X,Z).
\end{equation}

The convolution formula for spinor Wightman functions is obtained from (\ref{9}), (\ref{17}) in the same way:

\begin{equation}
\label{19}
\int_{dS} dY W^{(\frac{1}{2})}_{\lambda_{\psi}}(X,Y) \, W^{(\frac{1}{2})}_{\lambda_{\chi}}(Y,Z) = L^{d + 1} \, \Gamma \left(\frac{1}{2} \pm i \lambda_{\psi}\right) \, \delta(\lambda_{\psi} - \lambda_{\chi}) \, W^{(\frac{1}{2})}_{\lambda_{\psi}}(X,Z).
\end{equation}
and from (\ref{7}), (\ref{12}) (\ref{17}) it follows:

\begin{eqnarray}
\label{20}
W^{(0)}_{\lambda_{\phi}}(X,X) = \frac{\Gamma(h)} { 2 \pi^{h} \Gamma(2h)} \, \, \Gamma (h \pm i \lambda_{\phi}), \qquad  \nonumber
\\
\\
{\rm Tr}[ W^{(\frac{1}{2})}_{\lambda_{\psi}}(X,X)] = {\it dim \gamma} \, \, \frac{\Gamma(h)} { 2 \pi^{h} \Gamma(2h)} \, \, \Gamma (h + \frac{1}{2}\pm i \lambda_{\psi}).
\nonumber
\end{eqnarray}

The repeatition, with account of (\ref{17}), (\ref{18}), (\ref{20}), step by step, of the procedure of previous section, which allowed to get the 
EAdS Kallen-Lehmann decomposition (\ref{5}), (\ref{6}) for "two scalar fields harmonic bubble" permits to obtain the similar decomposition for the 
product of two scalar Wightman functions (the formulas below do not include poles from the complementary series in case the light scalar fields are considered; but in any way they may be applied for both heavy and light scalar fields, see comments in the end of the Introduction):

\begin{equation}
\label{21}
W^{(0)}_{\lambda_{1}}(X,Y) \, W^{(0)}_{\lambda_{2}}(X,Y) = \int^{+\infty}_{- \infty} dc \,  c \, \rho^{(0,0)}_{\lambda_{1}, \lambda_{2}} (c) \, W^{(0)}_{c}(X,Y),
\end{equation}
where

\begin{eqnarray}
\label{22}
\rho^{(0,0)}_{\lambda_{1}, \lambda_{2}} (\lambda_{3}) = \frac{1}{(2\pi)^{2} \, |\lambda_{3}|} \cdot \frac{\int_{dS} dX W^{(0)}_{\lambda_{1}}(X,Y) \, W^{(0)}_{\lambda_{2}}(X,Y) \, W^{(0)}_{\lambda_{3}}(X,Y)}{L^{d + 1} \, \, \, \Gamma (\pm i \lambda_{3}) \, \, \,  W^{(0)}_{\lambda_{3}}(X,X)} = \qquad \nonumber
\\  \nonumber
\\
= \frac{1}{(2\pi)^{2} \, |\lambda_{3}|} \, \, B^{(0,0)}_{i \lambda_{1}, i \lambda_{2}} (i \lambda_{3})  \cdot \frac{\Gamma (\pm i \lambda_{1}) \Gamma (\pm i \lambda_{2}) }{\Gamma(\pm i \lambda_{3})} = \qquad  \qquad  \qquad 
\\  \nonumber
\\ \nonumber
= \frac{1}{2^{5} \pi^{h+2}\Gamma (h)} \, \, \frac{1}{|\lambda_{3}|} \, \, \, \, \, \frac{\Pi_{\{\sigma_{i} = \pm\}} \Gamma\left( \frac{h + \sigma_{1} i \lambda_{1} + \sigma_{2} i \lambda_{2} + \sigma_{3} i \lambda_{3} }{2}\right)}{\Gamma(h \pm i \lambda_{3}) \, \, \Gamma ( \pm i \lambda_{3})},  \qquad  \qquad  \qquad  \nonumber
\end{eqnarray}
$B^{(0,0)}_{i \lambda_{1}, i \lambda_{2}} (i \lambda_{3})$ see in (\ref{6}), product $c \, \rho^{(0,0)}_{\lambda_{1}, \lambda_{2}} (c)$ is 
even under change of sign $c \to - c$, thus $\int_{-\infty}^{\infty} dc \, c \, \rho (c)$ in (\ref{21}) may be written as $\int_{0}^{\infty} dc^{2} \rho$;
this symmetry also allowed us to change $c \, \rho^{(0,0)}_{\lambda_{1}, \lambda_{2}} (c) \to |c| \, \rho^{(0,0)}_{\lambda_{1}, \lambda_{2}} (|c|)$ to have
 positive-definite Kallen-Lehmann density (\ref{22}). The introduction in denominator of (\ref{22}) of the seemingly artificial coefficient $(2\pi)^{2}$ is due to the choice of the normalisation of spectral representation of 
the AdS scalar Green functions in \cite{pened1011}, wherefrom coefficient in (\ref{3}) was taken.

Wonderful expression for $\rho^{(0,0)}_{\lambda_{1}, \lambda_{2}} (\lambda_{3} )$ in the third line of (\ref{22}) was first derived in \cite{Eps06} - \cite{Eps12} 
directly in dS space-time without "support reference" to AdS space. $\rho^{(0,0)}_{\lambda_{1}, \lambda_{2}} (\lambda_{3} )$ is 
introduced in (\ref{21}) following these papers in such a way that it comes in a flat space limit to the well-known expression for 
the Kallen-Lehmann density $\rho^{(0,0)}_{flat}$ of a one-loop diagram formed by two scalar fields of masses $M_{1}$ 
and $M_{2}$ \cite{Peskin}: 

\begin{eqnarray}
\label{23}
\int \, e^{ipx} \, D_{M_{1}} \, D_{M_{2}} \, d^{d+1} x = \int \frac{d^{d+1} k}{(2\pi)^{d+1}} \, \frac{1}{(k^{2} - M_{1}^{2} - 
i \epsilon)((p - k)^{2} - M_{2}^{2} - i \epsilon)} =   \nonumber
\\  \nonumber
\\
\\  \nonumber
= \int_{(M_{1} + M_{2})^{2}}^{\infty} \frac{d q^{2}}{q^{2} - p^{2} - i \epsilon} \, \rho^{(0,0)}_{flat} (q, M_{1}, M_{2}) \qquad \qquad \qquad  \qquad \, \,  
\end{eqnarray}
($D_{M}$ is $D_{M}(x - y)$ - Green function of scalar field of mass $M$ in $d + 1$ dimensional Minkowski space-time),

\begin{eqnarray}
\label{24}
\rho^{(0,0)}_{flat} (q, M_{1}, M_{2} ) =  \frac{1}{2^{4h - 1} \pi^{h} \Gamma(h)} \, \, \frac {\{ F(q, M_{1}, M_{2})\}^{h - 1} \, \, \theta [q - (M_{1} + M_{2})]  }{|q|^{2h - 1}} \nonumber
\\
\\
F(q, M_{1}, M_{2}) = [ q^{2} - (M_{1} + M_{2})^{2}]  \, \, [q^{2} - (M_{1} - M_{2})^{2}]. \nonumber
\end{eqnarray}
For $h = 3/2$ this expression for $\rho^{(0,0)}_{flat} (q)$ is the most familiar dimensionless Kallen-Lehmann density in four dimensions. 
(Integrals on the RHS of (\ref{23}) are UV-divergent in four dimensions; writing them in an arbitrary dimension ensures their dimensional regularization. Of course, it is necessary to add here the counterterms known from textbooks; we do not write them out explicitly, since the subject of our attention is the  Kallen-Lehmann density, the form of which does not depend on these refinements.)

Flat space Kallen-Lehmann density (\ref{24}) immediately follows from dS one (\ref{22}) in the limit $L \to \infty$ if in the last line 
of (\ref{22}) to 
substitute (see (\ref{16}) for $ML \gg h$) $\lambda_{1} = M_{1} L$, $\lambda_{2} = M_{2} L$, $\lambda_{3} = q L$ and to use the approximation  

\begin{equation}
\label{25}
\Gamma (a \pm i L b) \to 2\pi \, |b|^{2a - 1} L^{2a - 1} \, e^{ - \pi L |b|} 
\end{equation}
 at $L \to \infty$. Step function $\theta [q - (M_{1} + M_{2})]$ appears from the exponent $e^{-\pi L A /2}$ where 

\begin{equation}
\label{26}
A = |q + M_{1} + M_{2}| + \, |q - M_{1} - M_{2}| + \, |q + M_{1} - M_{2}| + \, |q - M_{1} + M_{2}| - 4 |q|
\end{equation}
is positive at $q < (M_{1} + M_{2})$ and is equal to zero at $ |q| > (M_{1} + M_{2})$. 
  
 Finally $\rho^{(0,0)}_{\lambda_{1}, \lambda_{2}} (\lambda_{3})$ of (\ref{22}) comes to $L^{d - 3} \rho^{(0,0)}_{flat} (q, M_{1}, M_{2}) $ of (\ref{24}) which 
 is correct result  from the point of view of dimensionalities, as it will be seen in the next Section.

Now let's turn to the main stated topic of this article - loops with spinors. Integral over dS space-time of the product of two spinor 
and one scalar Wightman functions is obtained from (\ref{4}), (\ref{17}) and (\ref{1}):

\begin{equation}
\label{27}
\int_{dS} dX {\rm Tr} [W^{(\frac{1}{2})}_{\lambda_{\psi}}(X,Y) \, W^{(\frac{1}{2})}_{\lambda_{\chi}}(X,Y)] \, W^{(0)}_{\lambda_{\phi}}(X,Y) = 
\frac{{\it dim \gamma}  \,  \, L_{AdS}^{d + 1}} {2^{5} \pi^{2h + 2} \Gamma(2h)}   \cdot     
{\cal {\bf N}}(\lambda_{\psi}, \lambda_{\chi}; \lambda_{\phi}),
\end{equation}
$ {\cal {\bf N}}(\lambda_{\psi}, \lambda_{\chi}; \lambda_{\phi})$ is the product of eight gamma-functions in nominator 
in (\ref{1}) where all $\nu$ in arguments are changed to the corresponding $i \lambda$ like in (\ref{16}):

\begin{eqnarray}
\label{28}
{\cal {\bf N}}( \lambda_{\psi},  \lambda_{\chi};  \lambda_{\phi}) = \,  \Gamma \left(\frac{h \pm i (\lambda_{\phi} + \lambda_{\psi} - 
\lambda_{\chi})}{2} \right) \, \, \Gamma \left(\frac{h \pm i (\lambda_{\phi} - \lambda_{\psi} + \lambda_{\chi})}{2} \right) \cdot \nonumber
\\
\\
\Gamma \left(\frac{1}{2}  + \frac{h \pm i ( \lambda_{\psi} + \lambda_{\chi} + \lambda_{\phi} )}{2} \right) \,  \, 
\Gamma \left(\frac{1}{2}  + \frac{h \pm i ( \lambda_{\psi} + \lambda_{\chi} - \lambda_{\phi} )}{2} \right). \qquad \nonumber
\end{eqnarray} 

Omitting intermidiate steps described above in the scalar case let's present, referring to (\ref{19}), (\ref{20}), (\ref{27}), (\ref{28}), the final expressions for the 
Kallen-Lehmann decompositions of product of two spinor Wightman functions (cf. (\ref{21}), (\ref{22}) of the 3-scalars case):

\begin{eqnarray}
\label{29}
{\rm Tr} [ W^{(\frac{1}{2})}_{\lambda_{\psi}}(X,Y) \, W^{(\frac{1}{2})}_{ - \lambda_{\chi}}(X,Y)] = \int^{+\infty}_{- \infty} dc \, c\, 
\rho^{(\frac{1}{2},\frac{1}{2})}_{\lambda_{\psi}, - \lambda_{\chi}} (c) \, W^{(0)}_{c}(X,Y), \qquad \nonumber
\\
\\
\rho^{(\frac{1}{2},\frac{1}{2})}_{\lambda_{\psi}, - \lambda_{\chi}} (c) =  \frac{1}{16 \pi^{h+2}\Gamma (h)} \, \, \frac{1}{|c|} \, \, \,  
\frac{{\cal {\bf N}}( \lambda_{\psi}, -  \lambda_{\chi};  c)}
{\Gamma(h \pm i c) \, \, \Gamma ( \pm i c)},  \qquad  \qquad  \qquad  \nonumber
\end{eqnarray}
and of product of spinor and scalar Wightman functions:

\begin{eqnarray}
\label{30}
W^{(\frac{1}{2})}_{\lambda_{\chi}}(X,Y) \, W^{(0)}_{ \lambda_{\phi}}(X,Y) = \int^{+\infty}_{- \infty} d{\bar c}\, 
\rho^{(\frac{1}{2},0)}_{\lambda_{\chi}, \lambda_{\phi}} ({\bar c}) \, W^{(\frac{1}{2})}_{{\bar c}}(X,Y), \qquad \nonumber
\\
\\
\rho^{(\frac{1}{2},0)}_{\lambda_{\chi}, \lambda_{\phi}} ({\bar c}) =  \frac{1}{16 \pi^{h+2}\Gamma (h)} \, \, \,  
\frac{{\cal {\bf N}}( \lambda_{\chi},  {\bar c},  \lambda_{\phi})}
{\Gamma(h + \frac{1}{2} \pm i {\bar c}) \, \, \Gamma (\frac{1}{2} \pm i {\bar c})}.  \qquad  \qquad  \qquad  \nonumber
\end{eqnarray}
The physical justification for the change of sign of $\lambda_{\chi}$ in expression (\ref{29}) for the two-spinors bubble was noted in Sec. 2, see comments after (\ref{14}).

Here ${\cal {\bf N}}(...)$ (\ref{28}) differs significantly from the similar product of eight 
gamma-functions in pure scalar case in numerator in the last line of (\ref{22}): now the arguments of four gamma functions in (\ref{28}) have additional terms 1/2. This gives the correct flat-space limit for 
the Kallen-Lehmann densities with participation of spinors \cite{Peskin}.

For two-spinors self-energy of scalar field in $d + 1$ dimensional Minkowski space-time we have (below $S_{m} = S_{m} (x - y)$ is Green function of spinor 
field of mass $m$,  $D_{M}$ is the same for scalar field of mass $M$):

\begin{eqnarray}
\label{31}
\int \, e^{ipx} \, {\rm Tr} [S_{m_{1}} \, S_{m_{2}} ]\, d^{d+1} x = \int \frac{d^{d+1} k}{(2\pi)^{d+1}} \, 
\frac{{\rm Tr} [ ({\hat k} + m_{1}) \, ({\hat p} - {\hat k} + m_{2})]}{(k^{2} - m_{1}^{2} - i \epsilon)((p - k)^{2} - m_{2}^{2} - i \epsilon)} = \nonumber
\\  \nonumber
\\
\\  \nonumber
= \int_{(m_{1} + m_{2})^{2}}^{\infty} \frac{d q^{2}}{q^{2} - p^{2} - i \epsilon} \, \rho^{(0,0)}_{flat} (q, m_{1}, m_{2} ) \cdot \frac{1}{2} \, 
 \, [q^{2} - (m_{1} -  m_{2})^{2}],  \quad \nonumber
\end{eqnarray}
$\rho^{(0,0)}_{flat} (q, m_{1}, m_{2} )$ see in (\ref{24}). 

Spectral integrals on the RHS of (\ref{29}), (\ref{30}) are UV-convergent. Regarding the UV-divergence in four dimensions of the flat space Kallen-Lehmann representations (\ref{31}) and (\ref{33}) below see comments after expression (\ref{24}).

Substitution in $\rho^{(\frac{1}{2},\frac{1}{2})}_{\lambda_{\psi}, - \lambda_{\chi}} (c)$ (\ref{29}) $\lambda_{\psi} = m_{1} L$, $\lambda_{\chi} = m_{2} L$, $c = q L$ gives 
at $L \to \infty$ with account of (\ref{28}), (\ref{25}), (\ref{26}) (where $M_{1,2}$ are changed to $m_{1,2}$) the wishful expression of (\ref{31}):

\begin{equation}
\label{32}
 \rho^{(\frac{1}{2},\frac{1}{2})}_{\lambda_{\psi}, - \lambda_{\chi}} (c) \to L^{d - 1} \, \, \rho^{(0,0)}_{flat} (q, m_{1}, m_{2} ) \cdot \frac{1}{2} \,  \, [q^{2} - (m_{1} -  m_{2})^{2}].
\end{equation}

The situation with the product of spinor and scalar Wightman functions is not so trivial. The familiar flat space expression for spinor 
one-loop self-energy formed by spinor of mass $m$ and scalar of mass $M$ looks as:

\begin{eqnarray}
\label{33}
\int \, e^{ipx} \, S_{m} \, D_{M} \, d^{d+1} x = \int \frac{d^{d+1} k}{(2\pi)^{d+1}} \, 
\frac{ ({\hat k} + m)}{(k^{2} - m^{2} - i \epsilon)((p - k)^{2} - M^{2} - i \epsilon)} =  \nonumber
\\  \nonumber
\\
\\  \nonumber
= \int_{(m + M)^{2}}^{\infty} \frac{d q^{2}}{q^{2} - p^{2} - i \epsilon} \, \rho^{(0,0)}_{flat} (q, m, M ) \cdot 
\left[ m + {\hat p} \, \, \frac{q^{2} + m^{2} - M^{2}}{2 q^{2}} \right]. \qquad   \qquad \, \,  
\end{eqnarray}
 
 At the same time for the $L \to \infty$ asymptotic of the Kallen=Lehmann density in (\ref{30}) after substitution $\lambda_{\chi} = m L$,
$\lambda_{\phi} = M L$, ${\bar c} = q L$ it is obtained:

\begin{equation}
\label{34}
 \rho^{(\frac{1}{2},0)}_{\lambda_{\chi}, \lambda_{\phi}} ({\bar c}) \to L^{d - 2} \, \, \rho^{(0,0)}_{flat} (q, m, M) \cdot 
 \frac{1}{2 |q|} \cdot f(q), \quad f(q) = |(q + m)^{2} - M^{2}|.
\end{equation}
With account of $|q| > m + M$ (we remind that this follows from the common factor $e^{- \pi L A /2}$ where $A(q, m, M)$ see in (\ref{26})) the 
odd and even parts of $f(q)$ look as:

\begin{equation}
\label{35}
 f(q) - f (-q)= 4 q m; \qquad  f(q) + f(- q) = 2 (q^{2} + m^{2} - M^{2}).
\end{equation}

To see the correspondence of (\ref{34})-(\ref{35}) and (\ref{33}) let us represent the flat space Kallen-Lehmann formula in the second line of (\ref{33}) in its 
"primordial" form with use of some density $\rho^{(\frac{1}{2}, 0)}_{flat|m, M}(q)$ which, generally speaking, is not symmetric under $q \to - q$:

\begin{eqnarray}
\label{36}
\int \, e^{ipx} \, S_{m} \, D_{M} \, d^{d+1} x = \int_{- \infty}^{\infty} dq \frac{q + {\hat p}}{q^{2} - p^{2} - i \epsilon} \,
 \, \rho^{(\frac{1}{2}, 0)}_{flat|m, M}(q) = \qquad  \qquad  \nonumber
\\ \nonumber
\\
\\ \nonumber
= \int_{0}^{\infty} \frac{dq}{q^{2} - p^{2} - i\epsilon} \, \, \left[q \cdot  \left(\rho^{(\frac{1}{2}, 0)}_{flat}(q) - \rho^{(\frac{1}{2}, 0)}_{flat}(- q) \right) +
 {\hat p} \cdot  \left(\rho^{(\frac{1}{2}, 0)}_{flat}(q) + \rho^{(\frac{1}{2}, 0)}_{flat}(- q) \right) \right]. \nonumber
\end{eqnarray}
Then it is immediately seen with account of (\ref{35}) that identification of the Kallen-Lehmann densities in (\ref{36}) and in (\ref{34}):

\begin{equation}
\label{37}
\rho^{(\frac{1}{2}, 0)}_{flat|m, M}(q) = \rho^{(0,0)}_{flat} (q, m, M) \cdot  \frac{1}{2 |q|} \cdot f(q)
\end{equation}
leads (\ref{36}) to form ({33}), that is, according to (\ref{34})

\begin{equation}
\label{38}
\rho^{(\frac{1}{2},0)}_{\lambda_{\chi}, \lambda_{\phi}} ({\bar c}) \to L^{d - 2} \, \, \rho^{(\frac{1}{2}, 0)}_{flat|m, M}(q).
\end{equation}

Kallen-Lehmann decompositions (\ref{29}), (\ref{30}) including explicit expressions for $\rho^{(\frac{1}{2},\frac{1}{2})}$ and $\rho^{(\frac{1}{2},0})$ are the main results of this section.

As noted in Sec. 2, the coefficient on the RHS of (\ref{4}) was written on the physical grounds, since it determined the coefficient in (27), and in turn expressions (29), (30) for Kallen-Lehmann densities $\rho^{(\frac{1}{2},\frac{1}{2})}$, $\rho^{(\frac{1}{2},0})$, which have the correct flat space limits thanks to the choice of coefficient in (\ref{4}). It would be interesting to test this choice by direct calculation.

Let us demonstrate how Kallen-Lehmann density $\rho^{(\frac{1}{2},0)}_{\lambda_{\chi}, \lambda_{\phi}} ({\bar c})$  (\ref{30}) looks like with specific example, where it is formed by a spinor of mass $m$ and a conformally invariant scalar field with $i\lambda_{\phi} = 1/2$ in all dimensions.

In this case nominator $N(...)$ (\ref{28}) in (\ref{30}) takes the form:

\begin{eqnarray}
\label{39}
{\cal {\bf N}} (mL, qL, i\lambda_{\phi} = 1/2) = \Gamma\left( \frac{h}{2} + \frac{1}{4} \pm i L \frac{q - m}{2}\right) \, \, 
\Gamma\left( \frac{h}{2} - \frac{1}{4} \pm i L \frac{q - m}{2}\right) \cdot \nonumber
\\
\\
\cdot \Gamma\left( \frac{h}{2} + \frac{3}{4} \pm i L \frac{q + m}{2}\right) \, \, \Gamma\left( \frac{h}{2} + \frac{1}{4} \pm i L \frac{q + m}{2}\right), \qquad \qquad  \nonumber
\end{eqnarray}
and for Kallen-Lehmann density (\ref{30}) in four dimensions, $h = 3/2$, it is obtained:

\begin{eqnarray}
\label{40}
\rho^{(\frac{1}{2},0)}_{m L, i \lambda_{\phi} = 1/2} ({q L}) =  \frac{L}{32 \pi^{2}} \, \,   \frac{[1 + L^{2} (q + m)^{2}] \, \, (q^{2} - m^{2})}
 {(1 + L^{2} q^{2}) } \cdot \nonumber
 \\
 \\
 \cdot \frac{\sinh ( 2\pi q L)}{q \, [\cosh(2 \pi q L) - \cosh(2 \pi m L)]}. \qquad  \qquad \qquad \qquad  \nonumber
\end{eqnarray}

Flat space limits of this expression, $L \to \infty$, gives the RHS of (\ref{38}), with account of the $M = 0$ versions of (\ref{24}), (\ref{34}), (\ref{37}) and the limit

\begin{equation}
\label{41}
 \frac{\sinh ( 2\pi q L)}{q \,  [\cosh(2 \pi q L) - \cosh(2 \pi m L)]} \to \frac{\theta (|q| - m)}{|q|}.
\end{equation}

Decay rate $1 / \tau$ ($\tau$ is the lifetime) of the particle entering the loop is proportional to Kallen-Lehmann densities (\ref{29}), (\ref{30}) which 
arguments $c$ or $\bar c$ are equalized to the conformal index of the entering particle \cite{Eps06}, \cite{bros0812}. From (40), as well as from general expressions (\ref{29}), (\ref{30}), it is seen the well known for dS space-time possibility 
of decay of entering particle of mass below the flat space threshold; the decay rate of such process is exponebtially suppressed what is clearly visible, for example, for $q < m$ in (40).

To conclude this Section let us consider the physically interesting case of light ($ML \ll 1$) or massless ($i\lambda_{\phi} = h$, see (\ref{16})) scalar field interacting with fermion of mass $m$. Late-time behavior of the heavy scalar field's one-loop self-energy formed with the same heavy field and light another scalar field was studied in \cite{Stout2310} where the observational consequences of such process during inflation were explored. Here we look at the similar diagram in frames of Yukawa model using the Kallen-Lehmann density $\rho^{(\frac{1}{2},0)}_{\lambda_{\chi}, \lambda_{\phi}} ({\bar c})$ (\ref{30}). In this case:

\begin{equation}
\label{42}
\lambda_{\chi} = mL; \quad i\lambda_{\phi} = \sqrt{h^{2} - M^{2} L^{2}} \approx h - \frac{M^{2}L^{2}}{2 h}; \quad \Delta_{\phi} = \frac{M^{2}L^{2}}{2 h},
\end{equation}
here of two possible values ($h \pm i\lambda_{\phi}$) of scalar field's conformal dimension $\Delta_{\phi}$ the smaller one responsible for the late-time divergences is chosen.

The asymptotic behavior of Wightman functions at  the dS future horizon, $\eta \to 0$ ($\eta$ is conformal time), is known, see \cite{Schaub} for spinors and \cite{Sleight1906}, \cite{Gorb2108}, \cite{Pened2306} for scalars:

\begin{equation}
\label{43}
W^{(0)}_{\lambda_{\phi}} (X, Y) \sim \frac{(\eta_{x}\eta_{y})^{\Delta_{\phi}}}{\Delta_{\phi}} \approx \frac{1}{\Delta_{\phi}} + \ln(\eta_{x}\eta_{y}); \quad W^{(\frac{1}{2})}_{\lambda_{\chi}} (X, Y)  \sim (\eta_{x}\eta_{y})^{\Delta_{\chi} + \frac{1}{2}},
\end{equation}
here $\Delta_{\phi} = M^{2}L^{2} / 2h \ll 1$; $\Delta_{\chi} = h + imL$. The $1/\Delta_{\phi} \sim 1/M^{2}$ term reflects the   standard zero-mode divergence of the massless scalar field propagators and must be subtracted with IR-counterterms, whereas the $\ln(\eta_{x}\eta_{y})$ term of the light scalar field's Wightman function is responsible for the late-time breakdown of the perturbation expansion typical for the light quantum scalar field on de Sitter space-time,
see \cite{Boyan1103} - \cite{Miao2405} and many other papers.

So, for the future asymptotics of product of two Wightman functions in the LHS of (\ref{30}) it is obtained from (\ref{42}) and (\ref{43}):

\begin{equation}
\label{44}
W^{(\frac{1}{2})}_{\lambda_{\chi}}(X,Y) \, W^{(0)}_{ \lambda_{\phi}}(X,Y) \sim \frac{(\eta_{x}\eta_{y})^{h + \frac{1}{2} + imL + \frac{(ML)^{2}}{2h}}}{(ML)^{2}/2h}.
\end{equation}
Let us show that the same IR-behavior is hidden in the Kallen-Lehmann  decomposition in the RHS of (\ref{30}).

In case $ML \ll 1$ it is possible to put $M = 0$ in the arguments of six from eight gamma-functions in nominator $N(...)$ (\ref{28}) in (\ref{30}) 
which then, with account of value of $i\lambda_{\phi}$ from (42), takes a form (the ubstitution ${\bar c} = q L$, $\lambda_{\chi} = m L$ is used here):

\begin{eqnarray}
\label{45}
{\cal {\bf N}} (mL, qL, \lambda_{\phi}) = \Gamma\left(\frac{(ML)^{2}}{4h} \pm i L \frac{q - m}{2}\right) \, \,  \Gamma\left( h \pm i L \frac{q - m}{2}\right)\cdot \nonumber
\\
\\
\cdot \Gamma\left( \frac{1}{2} + h \pm i L \frac{q + m}{2}\right) \, \, \Gamma\left( \frac{1}{2} \pm i L \frac{q + m}{2}\right).  \nonumber
\end{eqnarray}
Finally, using here for the first factor in the RHS the general formula

\begin{equation}
\label{46}
\Gamma(x \pm iy) = \Gamma(x)^{2} \, \Pi_{k=0}^{\infty} \frac{(x + k)^{2}}{y^{2} + (x + k)^{2}} \approx \frac{1}{x^{2} + y^{2}} 
\cdot \frac{\pi y}{\sinh \pi y} \quad (for \, \,  x \ll 1),
\end{equation}
for Kallen-Lehmann density (\ref{30}) in four dimensions and for $i\lambda_{\phi}$ (42) it is obtained:

\begin{eqnarray}
\label{47}
\rho^{(\frac{1}{2},0)}_{m L, \lambda_{\phi}} ({q L}) =  \frac{L}{32 \pi^{2}} \, \,   \frac{ [4 + L^{2} (q + m)^{2}] \, \, (q^{2} - m^{2})}{(1 + L^{2} q^{2})}   \cdot \frac{[1 + L^{2} (q - m)^{2}]}{[M^{4}L^{4}/4h^{2} + L^{2} (q - m)^{2}] } \cdot   \nonumber
 \\
 \\
\cdot  \frac{\sinh ( 2\pi q L)}{q \, [\cosh(2 \pi q L) - \cosh(2 \pi m L)]}. \qquad  \qquad \qquad \qquad  \nonumber
\end{eqnarray}

This KL density has poles at $qL = q_{*} L = m L \pm i (M^{2}L^{2}/ 2h)$. The leading late-time behavior of the Kallen-Lehmann decomposition comes from poles in KL density with minimal real part of value of conformal dimension, whereas spectral integral is given by the residues in these poles, see e.g. (3.26) in \cite{Pened2107}. Thus, using in the RHS of (\ref{30}) late-time asymptotic (43) of spinor Wightman function $W^{(\frac{1}{2})}_{qL}(X, Y)$ and making the correct choice of the closing contour in the complex q-plane gives a physically meaningful result for the RHS of (\ref{30}) which is proportional to:

\begin{equation}
\label{48}
\int^{+\infty}_{- \infty} d(qL) \, \rho^{(\frac{1}{2},0)}_{mL, \lambda_{\phi}} \, \,  (\eta_{x}\eta_{y})^{h + \frac{1}{2} + iqL}   \sim  
\frac{(\eta_{x}\eta_{y})^{h + \frac{1}{2} + imL + \frac{(ML)^{2}}{2h}}}{(ML)^{2}/2h}.
\end{equation}

This, as it could be expected, coincides with the asymptotic (44) of the LHS of KL decomposition (\ref{30}). We'll come back to the discussion of the KL density (\ref{47}) in the end of Sec. 4.

\section{Kallen-Lehmann representation of one-loop self-energies on dS. Spectral equations in "chain" approximation}

\qquad In contrast to the Kallen-Lehmann decomposition of the product of two Wightman functions considered in the previous Section, constructing a KL representation for the self-energy of a quantum field, which is the product of two time-ordered Feynman functions, is a more complex task. Below, we derive a KL representation on de Sitter space for the real part of the product of two scalar fields' Feynman Green functions, and write out the spectral equations accepting a hypothesis that the KL representation of any, scalar or spinor, one-loop diagrams on de Sitter space has a structure similar to analogous representations in flat space.

So, let's warm up on the 3-scalars theory on EAdS, $L_{int} = g \, \phi_{1} \phi_{2} \phi_{3}$, dimensionality of coupling 
constant $[g] = cm^{(d - 5)/2}$. Harmonic decomposition for the one-loop self-energy of field $\phi_{3}$ formed of Feynman Green 
functions of fields $\phi_{1} \, \phi_{2}$ looks as \cite{pened1011}:

\begin{equation}
\label{49}
g^{2} \, G^{AdS}_{\nu_{1}}(X, Y)\, G^{AdS}_{\nu_{2}}(X, Y) = \frac{g^{2}}{L_{AdS}^{2(d-1)}}  \, \int q_{\nu_{1}, \nu_{2}} (c) \,  \Omega^{AdS}_{c} (X, Y) \, dc,
\end{equation}
where for the harmonic decomposition of single Green function of positive $\nu$ it holds:

\begin{eqnarray}
\label{50}
G^{AdS}_{\nu}(X, Y)  = \frac{1}{L_{AdS}^{d - 1}} \, \int_{-\infty}^{\infty} \frac{dc}{c^{2} + \nu^{2}} \Omega^{AdS}_{c} (X, Y), \quad \nonumber
\\
\\
G^{AdS}_{ - \nu}(X, Y)  = G^{AdS}_{\nu}(X, Y) + \frac{2\pi i}{\nu} \, \, \Omega^{AdS}_{\nu} (X, Y). \quad  \nonumber
\end{eqnarray}
In \cite{pened1011} "harmonic image" $q_{\nu_{1}, \nu_{2}} (c)$ of product (\ref{49}) of two scalar Green functions 
is written down as a product of two spectral integrals (\ref{50}) with use of the harmonic decomposition (\ref{5}), (\ref{6}) of 
product of two harmonic functions. The "harmonic images" $q_{\nu_{1}, - \nu_{2}}$, $q_{- \nu_{1}, - \nu_{2}}$ of the 
products $ G^{AdS}_{\nu_{1}}(X, Y)\, G^{AdS}_{- \nu_{2}}(X, Y)$ and $ G^{AdS}_{-\nu_{1}}(X, Y)\, G^{AdS}_{-\nu_{2}}(X, Y)$ are obtained using 
expression (\ref{50}) of $G^{AdS}_{-\nu}$ through $G^{AdS}_{\nu}$. 

The named double spectral integral of \cite{pened1011} is UV-divergent in four dimensions (like its familiar flat space limit non-trivially obtained in \cite{pened1011}). Thus $q_{\nu_{1}, \nu_{2}} (c)$ in (\ref{49}) is infinite in four or higher integer dimensions, that is it needs regularization, and to get the physically meaningful self-energy the counterterms must be conventionally added to the bare Lagrangian. Another, actually rather artificial, way of the ”double-trace from UV to IR flow” subtraction of UV-infinities in calculation of self-energies was applied in \cite{alt-1}, \cite{alt-2}. In the present papere, when formally writing UV-divergent spectral expressions below, we always have in mind one or another procedure of regularization and subtraction of infinities.

Finally, with account of convolution formula (\ref{8}) and normalization condition

\begin{equation}
\label{51}
\frac{1}{L_{AdS}^{d + 1}}\, \int_{-\infty}^{\infty} d\nu \, \Omega^{AdS}_{\nu} (X, Y) = \frac{\delta^{(d+1)} (X, Y)}{\sqrt{- g^{(d + 1)}}},
\end{equation}
the harmonic decomposition of the full scalar Green function ${\bf G}^{AdS} (X, Y)$ obeying the inhomogeneous 
(with $\delta^{(d+1)} (X, Y)$ in the RHS) Dyson equation in chain approximation is obtained:

\begin{equation}
\label{52}
{\bf G}^{AdS} (X, Y) = \frac{1}{L_{AdS}^{d - 1}}\, \int_{-\infty}^{\infty} \frac{d\nu }
{\nu^{2} - \nu_{3\, (0)}^{2} - \frac{g^{2}}{L_{AdS}^{d-5}} \, q_{\nu_{1}, \nu_{2}} (\nu)} \, \,   \Omega^{AdS}_{\nu} (X, Y),
\end{equation}
here $\nu_{3 \, (0)}^{2} = h^{2} + L_{AdS}^{2}M_{3(0)}^{2}$, $M_{3(0)}$ is bare mass of the incoming field $\phi_{3}$, see (\ref{2}). 
Setting the denominator in (\ref{52}) equal to zero yields spectral equation for the scaling dimensions of the bound states in a given channel in chain approximation.

The application of the described AdS approach to de Sitter space-time is not an easy task. Formal substitution (\ref{16})
$\nu_{k} \to i \lambda_{k}$ in denominator of (\ref{52}) gives the spectral equation

\begin{equation}
\label{53}
\lambda^{2} - \lambda_{\phi_{3}\, (0)}^{2} - \frac{g^{2}}{L^{d-5}} \, {\rm Re} \left[{\hat \Pi}^{(0,0)}_{\lambda_{\phi_{1}} \lambda_{\phi_{2}}} (\lambda)\right] = 0,
\end{equation}
the real part of the self-energy "harmonic image" ${\hat \Pi}^{(0,0)}$ (which is defined below in (\ref{61})) is written down here since it determines 
the spectrum of bound states \cite{Gorb2108}. Also the obtained below Kallen-Lehmann representation is valid for the real part of the one-loop self-energy.

Formal substitution $\nu \to i\lambda$ in spectral equation (\ref{50}) will not give the spectral decomposition for the dS Green function 
because of the divergencies at the lightcone in dS space-time \cite{Gorb2108}. Nevertheless, such a substitution in (\ref{50}) allows us to 
obtain a simple expression for the real part of $ G^{AdS}_{\pm i\lambda}$ in a form of spectral integral in the sense of the principal value:

\begin{equation}
\label{54}
 {\rm Re} [ G^{AdS}_{i\lambda}] = {\rm Re} [ G^{AdS}_{ - i\lambda}] = 
 \frac{1}{L_{AdS}^{d - 1}} \, {\rm P} \int_{-\infty}^{\infty} \frac{dc}{c^{2} - \lambda^{2}} \Omega^{AdS}_{c} (X, Y),
\end{equation}

The non-trivial relationships between the bulk propagators and the correlators in EAdS and dS are considered in detail in \cite{Sleight1906} - \cite{Gorb2108}. One and the same EAdS Green function being analytically continued to dS gives rise to four functions $G^{\alpha\beta} (X, Y)$,
$\alpha, \beta = l, r$ which correspond to left and right parts of Keldysh contour that is to different ways of regulating the singularity on
the lightcone:

\begin{equation}
\label{55}
G^{ll}(X, Y) = \langle 0 |T[\phi(X) \phi(Y)] | 0 \rangle; \quad G^{rr}(X, Y) = \langle 0 | {\bar T} [\phi(X) \phi(Y)] | 0 \rangle;
\end{equation}
whereas $G^{lr}(X, Y) = G^{rl}(X, Y) = W^{(0)}(X, Y) / L^{d-1}$, dimensionless Wightman function $W^{0}(X, Y)$ is introduced in Sec. 3, see (\ref{17}).

In \cite{Taron2109}, \cite{Gorb2108} it is shown that dS propagators $G^{\alpha\beta} (X, Y)$ can be analytically continued to a linear combination
of the AdS propagators $G^{AdS}_{\pm i\lambda}$. For example for the time-ordered dS Green function it was obtained:

\begin{equation}
\label{56}
G^{ll}_{\lambda}(s^{dS}) \to - \left(\frac{L_{AdS}}{L}\right)^{d-1}\, \frac{\lambda}{2\pi} \, \Gamma(\pm i\lambda)\, \, 
\left( G^{AdS}_{i\lambda} (s^{AdS})  \, e^{-\lambda} - G^{AdS}_{- i\lambda} (s^{AdS})  \, e^{\lambda}\right),
\end{equation}
$s^{dS}$, $s^{AdS}$ are invariant distances berween points $X, Y$ in dS and in EAdS  correspondingly; 
substitution $L_{AdS} = + i L$ (we remind that $L = L_{dS}$ throughout this paper) returns (56) to formula (3.19) in \cite{Gorb2108}.

Now let us pay attention that the substitution into (\ref{56}) of $G^{AdS}_{-i\lambda}$ from the second line of (\ref{50}) reduces, 
with account that 

$$
\lambda \, \sinh \pi\lambda \, \Gamma(\pm i\lambda) / \pi = 1,
$$
the RHS of (\ref{56}) to 

\begin{equation}
\label{57}
G^{ll}_{\lambda}(s^{dS}) \to - \left(\frac{L_{AdS}}{L}\right)^{d-1}\,  G^{AdS}_{i\lambda} (s^{AdS}) + H.T.,
\end{equation}
where $H.T.$ are homogeneous terms proportional to $\Omega^{AdS}_{i\lambda}$ that is obeing the homogeneous Klein-Gordon equation. 

Finally from (\ref{54}) and (\ref{57}) it follows the looked for harmonic decomposition of the real part of the time ordered dS Green function:

\begin{equation}
\label{58}
{\rm Re} [ G^{ll}_{\lambda} (X, Y)] = {\rm Re} [ G^{ll}_{ - \lambda} (X, Y)] \to 
 \frac{1}{L^{d - 1}} \, {\rm P} \int_{-\infty}^{\infty} \frac{dc}{\lambda^{2} - c^{2}} \Omega^{AdS}_{c} (X, Y).
\end{equation}

We need similar harmonic decomposition for the one-loop self-energy "bubble" which is the product $G^{ll}_{\lambda_{1}}(s^{dS}) \, G^{ll}_{\lambda_{2}}(s^{dS})$ of two Green function (56). In \cite{Marolf}, \cite{Taron2109}, \cite{Gorb2108} the harmonic images of such bubbles were  investigated, and quite complicated expressions for these images were obtained. In \cite{Taron2109}, \cite{Gorb2108} they are presented as a linear combination of the corresponding EAdS harmonic images $q_{\nu,\nu}$, $q_{\nu, -\nu}$, $q_{- \nu, -\nu}$ which by itself are given by rather complicated expressions. Here essentially more simple expressions for the harmonic decompositions of the products of two dS Green functions are proposed.

Perhaps, the most transparent way to study dS propagators is to look at the analytical continuation from sphere $S^{D}$ to
the D-dimensional de Sitter space-time \cite{Marolf}, \cite{Hollands}, \cite{Taron1907}, \cite{Gorb2108}. In Appendix C of \cite{Gorb2108}, the spectral representation is derived in this way for general two-point function $F^{\alpha \beta}(X, Y) = \langle 0 | O(X^{\alpha})\, O(Y^{\beta}) | 0 \rangle$ 
($\alpha, \beta = l, r$, we again omit in (\ref{59})  the possible contribution of poles of light scalar fields from the complementary series, see comments in the end of the Introduction):

\begin{equation}
\label{59}
F^{\alpha\beta}(X, Y) = \int_{- \infty}^{\infty} dc \, \frac{c}{\pi i} \, f(c) \, G_{c}^{\alpha \beta}(X, Y).
\end{equation}
In our case $O(X^{\alpha}) = \phi_{1} (X^{\alpha}) \phi_{2}(X^{\alpha})$, then $F^{lr}$ is proportional to the product 
$W^{(0)}_{\lambda_{1}}(X,Y) \, W^{(0)}_{\lambda_{2}}(X,Y)$ of two  Wightman functions of Sec. 3. And (\ref{59}) is the 
familiar Kallen-Lehmann decomposition (\ref{21}) of this product. Because of the symmetry of Wightman function, $G^{lr}_{c} = G^{lr}_{- c}$, 
comparison of (\ref{59}) with (\ref{21}) permits to identify:

\begin{equation}
\label{60}
\frac{c}{\pi i} \, \frac{f(c) - f(-c)}{2} = \frac{1}{L^{d-1}}  \, c \, \rho^{(0,0)}_{\lambda_{1}, \lambda_{2}} (c).
\end{equation}

Taking into account that real part of $G^{ll}_{c} (X, Y)$ in the RHS of (\ref{59}) is also, like Wightman function, symmetric ander $c \to - c$
(see (\ref{58})), it is seen that real part of product of two time-ordered Green 
functions ${\rm Re} F^{ll} (X, Y) = {\rm Re} [G^{ll}_{\lambda_{1}} (X, Y)\, G^{ll}_{\lambda_{2}} (X, Y)]$ is expressed in (\ref{59}) through the same
Kallen-Lehmann density $\rho^{(0,0)}_{\lambda_{1}, \lambda_{2}}$. 

Thus, after analytical continuation to EAdS in the RHS of (\ref{59}) and using spectral integral (\ref{58})
(where the variables $c$ and $\lambda$ are swapped)  the desired Kallen-Lehmann representation of the real part of the product of two
 time-ordered dS scalar Green functions is obtained:
 
 \begin{eqnarray}
\label{61}
{\rm Re} [G^{ll}_{\lambda_{1}} (X, Y)\, G^{ll}_{\lambda_{2}} (X, Y)] \to \frac{1}{L^{2(d-1)}} \, 
 \int_{-\infty}^{\infty}  {\rm Re} \left[{\hat \Pi}^{(0,0)}_{\lambda_{\phi_{1}} \lambda_{\phi_{2}}} (\lambda)\right] \, \Omega_{i \lambda}^{AdS}(X, Y) \, d\lambda = \nonumber
\\
\\
= \frac{1}{L^{2(d-1)}} \, \int_{-\infty}^{\infty}  \left[{\rm P} \int_{0}^{\infty} \frac{dc^{2}}{c^{2} - 
\lambda^{2}} \, \rho^{(0,0)}_{\lambda_{1}, \lambda_{2}} (c) \right] \, \, \Omega_{i \lambda}^{AdS}(X, Y) d\lambda. \qquad \nonumber
\end{eqnarray}

Substitution of the expression in square brackets in (\ref{61}) into (\ref{53}) gives the following rather simle spectral equation for conformal dimension $\lambda$ in the 3-scalars theory on dS in chain approximation:

\begin{equation}
\label{62}
\lambda^{2} - \lambda_{\phi_{3}\, (0)}^{2} - \frac{g^{2}}{L^{d-5}} \, {\rm Re} \left[{\rm P} \int_{0}^{\infty} \frac{dc^{2}}{c^{2} - \lambda^{2}} \, \rho^{(0,0)}_{\lambda_{1}, \lambda_{2}} (c) \right] = 0.
\end{equation}
We guess that this spectral equation (after standard procedures for accounting for UV-divergences) is suitable for practical calculations since Kallen-Lehmann density in (\ref{62}) is given by simple and compact formula (\ref{22}).

The familiar flat space limit of spectral equation (\ref{62}) is obtained, like in Sec. 3, with the substitutions 
$\lambda \to p L = M_{b} L$ ($p^{2} = M_{b}^{2}$, $M_{b}$ is mass of the bound state), $c \to q L$, $\lambda_{\phi_{1}} \to M_{1} L$, $\lambda_{\phi_{2}} \to M_{2} L$, $\lambda_{\phi_{3} (0)} \to M_{3 (0)} L$,
and with account that $\rho^{(0,0)}_{\lambda_{1}, \lambda_{2}} (c) $ (\ref{22}) $\to$ $L^{d-3} \, \rho^{(0,0)}_{flat} (q, M_{1}, M_{2} )$ (\ref{24}):

\begin{equation}
\label{63}
M_{b}^{2} - M_{3 (0)}^{2} - g^{2} \, {\rm P} \int_{0}^{\infty} \frac{dq^{2} }{q^{2} - M_{b}^{2}} \, \rho^{(0,0)}_{flat} (q, M_{1}, M_{2} ) = 0.
\end{equation}

The logic of deriving similar spectral equations in the Yukawa model ($L_{int} = e {\bar\psi}\psi \phi$, dimensionality of coupling 
constant $[e] = cm^{(d - 3)/2}$) is the same, as it may be supposed.
We will not repeat ourselves, but put forward the hypothesis that the spectral equations for conformal indices in the Yukawa model 
on de Sitter space-time have a form similar to the corresponding spectral equations in Minkowski space, like it took place in the model of three 
interacting scalars considered above.

Thus, it is propsed the following spectral equation for scaling index $\lambda_{\phi b}$ of scalar field of bare 
mass $M_{(0)}$ with its self-energy formed by the loop of two spinors of masses $m_{1}$, $m_{2}$:

\begin{equation}
\label{64}
\lambda_{\phi b}^{2} - \lambda_{\phi\, (0)}^{2} - \frac{e^{2}}{L^{d-3}} \, {\rm P} \int_{0}^{\infty} \frac{dc^{2}}{c^{2} - \lambda_{\phi b}^{2}} \, \rho^{(\frac{1}{2},\frac{1}{2})}_{\lambda_{\psi}, - \lambda_{\chi}} (c) = 0,
\end{equation}
$\lambda_{\phi b}^{2} = M_{b}^{2} L^{2} - h^{2}$, $\lambda_{\phi \, (0)}^{2} = M_{0}^{2} L^{2} - h^{2}$, $\lambda_{\psi} = m_{1} L$, $\lambda_{\chi} = m_{2} L$, $\rho^{(\frac{1}{2},\frac{1}{2})}_{\lambda_{\psi}, - \lambda_{\chi}} (c)$ see in (\ref{29}). Flat space limit of this equation is obtained with account of 
the flat space limit (\ref{32}) of the Kallen-Lehmann density $\rho^{(\frac{1}{2},\frac{1}{2})}$:

\begin{equation}
\label{65}
M_{b}^{2} - M_{(0)}^{2} - e^{2} \, {\rm P} \int_{0}^{\infty} \frac{dq^{2} }{q^{2} - M_{b}^{2}} \, 
\rho^{(0,0)}_{flat} (q, m_{1}, m_{2} ) \, \cdot \frac{1}{2} \,  \, [q^{2} - (m_{1} -  m_{2})^{2}] = 0.
\end{equation}

And for the spectral equation for the scaling index $\lambda_{\psi b} = m_{b} L$ of the bound state spinor field $\psi_{b}$ of 
bare mass $m_{0}$ ($\lambda_{\psi \, (0)} = m_{0} L )$, which one-loop self-energy is formed 
by spinor $\chi$ of mass $m$  ($\lambda_{\chi} = m L$) and scalar $\phi$ of mass $M$ ($\lambda_{\phi} = \sqrt{M^{2} L^{2} - h^{2}}$) it is proposed
to reproduce the flat space form of spectral integral given in the first line of (\ref{36}) (where substitution ${\hat p} \to m_{b}$ was performed, having in mund the Dirac equation for the bound state spinor field $({\hat p} - m_{b}) \psi_{b} = 0$):

\begin{equation}
\label{66}
\lambda_{\psi b} - \lambda_{\psi\, (0)} - \frac{e^{2}}{L^{d-3}} \, {\rm P} \int_{-\infty}^{\infty} \, d{\bar c}\frac{{\bar c} + \lambda_{\psi b}}
{{\bar c}^{2} - \lambda_{\psi b}^{2}} \, \rho^{(\frac{1}{2},0)}_{\lambda_{\chi}, \lambda_{\phi}} ({\bar c}) = 0,
\end{equation}
$\rho^{(\frac{1}{2},0)}_{\lambda_{\chi}, \lambda_{\phi}} ({\bar c})$  is given in (\ref{30}).

The flat space limit, $L \to \infty$, of (\ref{66}) is obtained with account of ${\bar c} = q L$, 
$\rho^{(\frac{1}{2},0)}_{\lambda_{\chi}, \lambda_{\phi}} ({\bar c}) \to L^{d - 2} \, \rho^{(\frac{1}{2}, 0)}_{flat | m, M}(q)$ (\ref{38}):

\begin{equation}
\label{67}
m_{b} - m_{0} - e^{2} \, {\rm P} \int_{-\infty}^{\infty} \, dq \, \frac{q + m_{b}}
{q^{2} - m_{b}^{2}} \, \rho^{(\frac{1}{2},0)}_{flat|m, M} (q) = 0.
\end{equation}
In both equations (\ref{66}), (\ref{67}) one can proceed to the integral $\int_{0}^{\infty} dq^{2}$ as it is done in (\ref{36}).

The spectral integral in (\ref{66}) is logarithmically UV-divergent for the Kallen-Lehmann density (\ref{30}) of the general form, for its special cases (\ref{40}), (\ref{47}) as well as in the flat space limit (\ref{67}). Thus  it needs hard-cut or dimensional regularization and  $\lambda_{\psi\, (0)}$
in (\ref{66}) must be supplemented with a counterterm in a standard way. 

Also, for the specific case of Yukawa model with light scalar field considered in the end of Sec. 3, it is seen from (\ref{47}) that for the massless scalar field, $M = 0$,
spectral integral in (\ref{66}) is IR-divergent at ${\bar c} = \lambda_{\chi}$ ($q = m$) for any value of $L < \infty$, but not in flat space. The degree of this IR-singularity increases further if the bootstrap condition $\lambda_{\psi_{b}} = \lambda_{\chi}$ ($m_{b} = m$) is imposed in the integrand in (\ref{66}) where KL density is taken from (\ref{47}). It is most likely that this paradoxical difference of spectral equations (\ref{66}) in flat space and in dS of any, arbitrarily small, curvature is due to the breakdown of the perturbation theory near the dS future horizon, that is in vicinity of the KL density pole, and that this paradox may be cured with the known resummation of the leading IR-contributions in all loops 
\cite{Boyan1103} - \cite{UGO1901}, \cite{loops2311}, \cite{Miao2405}, \cite{Instab2507} and references therein. But studying these issues is a topic for another paper.

\section{Conclusion}

\qquad This article is mainly of a technical nature. Its results are compact formulas (\ref{1}) and (\ref{28}) through which the
 Kallen-Lehmann decompositions for products of two spinors and of spinor and scalar harmonic functions in 
EAdS ((\ref{10}), (\ref{11}) and (\ref{13}), (\ref{14})), and for similar products of Wightman functions in dS ((\ref{29}) and (\ref{30})) are expressed in frames of Yukawan model. The first, also technical, task for the future could be probably the derivation of formulas like (1), (28) for fermions on EAdS and dS interacting with gauge fields, as it takes place in the physically more interesting QED and Standard Model, and also with graviton..

Kallen-Lehmann approach is a powerful tool potentially more efficient and simple than space-time calculations of quantum diagrams. For example, in \cite{alt-1}, \cite{alt-2} KL representations of the one-loop diagrams based on formulas (\ref{1}) and (\ref{3}) made it possible to write down non-trivial spectral equations for the conformal dimensions of scalar and spinor fields on the EAdS space and calculate the corresponding spectra.
In Sec. 4 these suitable for calculations EAdS spectral equations are giveneralized to the physically more significant de Sitter space-time. This generalization is based on the hypothesis put forward in the present paper that the structure of the KL representation of the one-loop self-energy on dS is similar to the structure of such a representation in flat space-time. The justification for this hypothesis presented in the paper for the 3-scalars model requires generalization to diagrams involving spinor lines.

The singularity of the KL density (47) in de Sitter space of any, arbitrarily small, curvature, but not in flat space, in the Yukawa model with a massless scalar field (discussed at the end of Sections 3 and 4) perhaps deserves a special study. Is it possible to formulate a resummation procedure, that could potentially cure this singularity, in the language of the Kallen-Lehmann representations ? Is the difference in the late-time behavior, visible in expression (\ref{47}) at $M = 0$, between quantum theories in de Sitter space and in flat space preserved for fermions interacting with graviton or with massless gauge field? These questions are not of the academic character, since, after all, we do not live in a flat space, but in a quasi-de Sitter space, albeit one of extremely small curvature, $1/L_{dS} = H = 10^{-32} eV$.

The results of this paper will hopefully permit to calculate  the one-loop corrections to the inflationary correlation functions with spinor lines, particle decay rates, diagrams with spinors with a larger number of loops, etc. They can thus be useful in a variety of vast areas of research related 
to inflation, including: 

- in the framework of the cosmological collider and cosmological bootstrap approaches, \cite{Ferm1612}, \cite{Arkani-1}, \cite{boot1} - \cite{boot5}; 

- in gravitational particle production and reheating after inflation which can play an important role in the cosmic history (dark matter, 
gravitational-wave radiation, the baryon asymmetry...), \cite{GPP2112} - \cite{GPP2410} and references therein; 

- in the controversial issue of stability/instability of (quasi-) de Sitter space-time, in the context of the infrared divergencies arising from 
loops in particular, see recent papers \cite{Akhm2411}, \cite{Instab2502} - \cite{Instab2507} and references therein.

Also, the results of this work can be hopefully useful in solving the problem posed in Conclusion of \cite{Arkani-2} to generalize 
to fermions and in general to the fields of the Standard Model the "All Loops" approach considered in \cite{Arkani-2} for bosons.

\section*{Acknowledgements} Author is grateful to Sergey Godunov, Anatoly Shabad, Alexander Smirnov, Boris Voronov and Mikhail Vysotsky for fruitful discussions and assistance, as well as to Ugo Moschella and Joao Penedones for useful comments on the preprint of the paper, and to JHEP reviewer for constructive criticism and important suggestions.

\end{document}